\documentclass[twocolumn]{aastex63}

\usepackage[T1]{fontenc}
\usepackage{amsmath}

\newcommand\won{\emph{W}1}
\newcommand\wtw{\emph{W}2}
\newcommand\wth{\emph{W}3}
\newcommand\wfo{\emph{W}4}
\newcommand{\wise}{\emph{WISE}}
\newcommand{\neowise}{\emph{NEOWISE-R}}
\newcommand{\wonetwo}{($W$1$-W$2)}

\shorttitle{Black Holes in Green Peas}
\shortauthors{Harish et al.}

\begin{document}

\title{\large Evidence for Black Holes in Green Peas from \emph{WISE} colors and variability}

\author{Santosh Harish}
\affil{School of Earth and Space Exploration, Arizona State University, Tempe, AZ 85287, USA}
\email{santosh.harish@asu.edu}

\author{Sangeeta Malhotra}
\affil{NASA Goddard Space Flight Center, 8800 Greenbelt Road, Greenbelt, MD 20771, USA}
\affil{School of Earth and Space Exploration, Arizona State University, Tempe, AZ 85287, USA}

\author{James E. Rhoads}
\affil{NASA Goddard Space Flight Center, 8800 Greenbelt Road, Greenbelt, MD 20771, USA}
\affil{School of Earth and Space Exploration, Arizona State University, Tempe, AZ 85287, USA}

\author{Tianxing Jiang}
\affil{School of Earth and Space Exploration, Arizona State University, Tempe, AZ 85287, USA}

\author{Huan Yang}
\affil{Las Campanas Observatory, Carnegie Institution for Science, Chile}

\author{Kendrick Knorr}
\affil{School of Earth and Space Exploration, Arizona State University, Tempe, AZ 85287, USA}



\begin{abstract}
We explore the presence of active galactic nuclei (AGN)/black holes (BH) in Green Pea galaxies (GPs), motivated by the presence of high ionization emission lines such as He\textsc{ii} and [Ne\textsc{iii}] in their optical spectra. In order to identify AGN candidates, we used mid-infrared (MIR) photometric observations from the all-sky Wide-field Infrared Survey Explorer (\wise) mission for a sample of 1004 GPs. Considering only $>5\sigma$ detections with no contamination from neighboring sources in AllWISE, we select 31 GPs out of 134 as candidate AGN based on a stringent 3-band \wise\ color diagnostic. Using multi-epoch photometry in \won\ and \wtw\ bands based on time-resolved unWISE coadd images, we find two sources exhibiting variability in both the \wise\ bands among 112 GPs with W1$\leqslant16$ mag and no contamination from neighboring sources in unWISE. These two variable sources were selected as AGN by the \wise\ 3-band color diagnostic as well. Compared to variable AGN fractions observed among low-mass galaxy samples in previous studies, we find a higher fraction ($\sim1.8\%$) of MIR variable sources among GPs, which demonstrates the uniqueness and importance of studying these extreme objects. Through this work, we demonstrate that MIR diagnostics are promising tools to select AGN that may be missed by other selection techniques (including optical emission-line ratios and X-ray emission) in star-formation dominated, low-mass, low-metallicity galaxies. 
\end{abstract}

\keywords{galaxies: dwarf -- galaxies: starburst -- galaxies: nuclei -- galaxies: evolution}


\section{Introduction} \label{sec:intro}
Since their discovery more than a decade ago, \emph{Green Peas} (GPs; \citealt{Cardamone2009}), a class of low-redshift ($z \lesssim 0.3$) extreme emission-line galaxies, have been studied extensively across the entire electromagnetic spectrum. These galaxies exhibit intense star formation activity, strong nebular lines, low metallicity, low dust content and high gas pressures (e.g., \citealt{Izotov2011,Yang2016,Yang2017a,Jiang2019,Kim2020}). They are also compact in size \citep{Kim2021} and typically contain low stellar mass. UV studies of Green Peas have shown the prevalence of Ly$\alpha$ emission-line with an equivalent width distribution matching those of high-redshift ($z \sim 3-6$) Ly$\alpha$ emitters (e.g., \citealt{Henry2015,Yang2016,Yang2017b}). They exhibit high [O\textsc{iii}]/[O\textsc{ii}] ratios, which are seen in some high-$z$ galaxies, and which may indicate the presence of optically thin ionized regions (e.g., \citealt{Jaskot2013, Nakajima2014}).  And, there is growing evidence that Lyman continuum radiation escapes more readily in Green Peas (with escape fractions 2-72\%; \citealt{Izotov2016, Yang2017b, Izotov2018}) than in any other known galaxy population, making them the best available low-$z$ analogs of the galaxies that drove cosmological reionization. GPs are selected on the basis of optical emission lines (generally [O\textsc{iii}]) with extreme equivalent widths. They often exhibit high ionization nebular lines such as He\textsc{ii} 4686\AA\ and [Ne\textsc{iii}] 3869\AA, which suggest a hard ionizing source such as active galactic nuclei (AGN).  However, the relative importance of star formation and accretion in powering GP activity remains poorly known.

Over the last few decades, there is a growing consensus that most massive galaxies host super-massive black holes (SMBHs) in their central regions \citep{Kormendy1995,Kormendy2013}. Less is known about black holes (BHs) among low-mass galaxies, although there has been an increase in low-mass BH candidates in the recent times (e.g., \citealt{Greene2007,Reines2013}). Low-mass galaxies can be crucial in placing constraints on models of BH seed formation and non-merger models of BH evolution \cite[e.g.,][]{Volonteri2009,Wassenhove2010,Greene2012}. Detection of AGN signatures in low-mass galaxies is a significant challenge. A large fraction of these AGNs may be missed by optical selection either because the central engine is obscured, or because emission-line indicators such as [O\textsc{iii}] are contaminated by extreme star formation in the host galaxy \citep[e.g.,][]{Goulding2009,Trump2015}. With X-ray selection, the presence of high-mass X-ray binaries (HMXBs) and ultra-luminous X-ray sources (ULXs) pose a serious challenge in the detection of low-mass BHs because of their comparable luminosities (e.g., \citealt{Lehmer2010,Mineo2012}).

Mid-infrared (MIR) AGN selection has shown promise in identifying some obscured, optically hidden AGN in several galaxies. The hard radiation field produced by the AGN can heat dust to high temperatures, generating strong MIR continuum and an infrared SED that is distinct from typical star-forming galaxies. The AGN continuum emission is approximately a power law in the $3-10\mu$m range, because strong UV and X-ray radiation destroys the molecules responsible for the Polycyclic Aromatic Hydrocarbon (PAH) emission, while heating the surrounding grains to near dust sublimation temperatures ($1000-1500$K). With the advent of Wide-field Infrared Survey Explorer (\wise), a large number of AGN have been identified, demonstrating that MIR color diagnostics can select luminous AGN with a 95\% reliability (e.g., \citealt{Stern2012}) 

An alternate way to identify AGN is based on flux variability. Active nuclei are known to be variable in different spectral regimes with timescales ranging from hours to years (e.g., \citealt{Sesar2007,Kozlowski2016}). The variability is thought to be related to thermal fluctuations in the accretion disk driven by a turbulent magnetic field \citep{Kelly2009} and/or temperature fluctuations in an inhomogenous accretion disk \citep{Ruan2014}. Variability-based selection is a promising means for finding massive BHs in dwarf galaxies. Several studies in the past have shown that such a selection can (1) identify AGN in the low-mass, low-metallicity regime missed by other selection techniques (e.g., \citealt{Baldassare2018}), (2) identify low-luminosity AGN \citep{Trevese1994}, and (3) identify AGN in dwarf galaxies where host galaxy emission dominates the total luminosity (e.g., \citealt{Trump2015,Secrest2020}).

In this paper, we present findings from the first-ever systematic search for AGN in GPs using MIR observations. Section \ref{sec:data} describes the GP sample and the relevant MIR \emph{WISE} photometric data. The MIR color and variability diagnostic used for AGN selection is presented in Section \ref{sec:agn_sel}. The results and comparison with other selection methods are discussed in Section \ref{sec:disc}. Our main conclusions are presented in Section \ref{sec:concl}. All magnitudes presented in this work are in Vega system, unless stated otherwise.

\section{Sample and Data} \label{sec:data}
The sample of Green Pea galaxies were selected following \citet{Jiang2019}. In total, there are 1004 objects that are spatially compact (R$_{90}$ $< 3\arcsec$), which contain well-detected (SNR$> 5$), strong [O\textsc{iii}]$_{\lambda5007}$ and/or H$\beta$ emission-lines (EW([O\textsc{iii}$_{\lambda5007}$]) $> 300\AA$ and/or EW(H$\beta$) $> 100\AA$), and have a literature spectral classification of ``Galaxy'' with a sub-class of ``starforming,''  ``starburst,'' or ``NULL.''   The literature classifications are drawn from the \emph{galSpecLine} \citep{Brinchmann2004,Tremonti2004} and \emph{emissionLinesPort} \citep{Maraston2013,Thomas2013} catalogs, based on SDSS Data Release 8 \citep{Aihara2011} and 12 \citep{Alam2015}, respectively. The object classification is based on the measured optical emission-line ratios.  However, as discussed in Section \ref{sec:bpt_cons}, some of these sources might still contain an AGN that is indistinguishable from a star-forming object based on optical spectroscopy alone.

\subsection{Mid-infrared data} \label{sec:wise_midIR}
The WISE mission \citep{Wright2010} conducted a mid-infrared imaging survey of the entire sky using 3.4, 4.6, 12, and 22 $\mu$m bandpasses (hereafter, \won, \wtw, \wth\ and \wfo), in 2010. With a field-of-view of $47\arcmin \times 47\arcmin$ and angular resolution of $\sim$6$\arcsec$ (\won, \wtw, \wth) and $\sim$ 12$\arcsec$ (\wfo), WISE scanned the entire sky once every six months with 12 or more independent, single-exposures of each point on the sky. The WISE solid hydrogen cryogen was exhausted in late 2010, and the spacecraft was put into hibernation in February 2011 after surveying the full sky twice. The NEOWISE-R Post-Cryo mission \citep{Mainzer2011} reactivated the spacecraft and has been surveying the sky from December 2013 till present using the two bluer bandpasses, \won\ and \wtw\ (which remain usable after cryogen exhaustion).

The AllWISE Source Catalog contains the most reliable and accurate MIR photometry for over 747 million objects on the sky based on observations from the WISE cryogenic and NEOWISE post-cryo phases of the mission. This catalog is derived from the AllWISE Atlas intensity images (1.37\arcsec/pixel) which consists of co-added 7.7s \won\ and \wtw\ single-exposure images from the 4-Band, 3-Band cryogenic and post-cryo mission phases, and co-added 8.8s \wth\ and \wfo\ single-exposure images from the 4-Band mission phase. In order to find the MIR counterparts of GPs, we cross-matched them, based on their SDSS coordinates, with the AllWISE Source Catalog using a search radius of 2\arcsec\ and found matches for 516 GPs from our sample. 

By stacking together single-exposure images from the same visit, the unWISE team \citep{Lang2014,Meisner2017a} has produced time-resolved coadds over a multi-year baseline \citep{Meisner2018} using \wise\ and \neowise\ observations. A typical sky location is imaged every six months by \wise\ with $\gtrsim12$ exposures per visit. In unWISE, these individual exposures (with $\sim1$ day intervals) are stacked to produce one coadd per band per visit, thereby creating one coadd every six months for a given position on the sky in each of the WISE bands, \won\ and \wtw. These coadds offer a powerful new tool for detecting variability of relatively faint \wise\ sources on long time-scales ($>0.5$ year) with detection of sources $\sim1.3$ mag fainter than the single-exposure depth \citep{Meisner2018}.

In order to select a reliable sample of variable sources, we restrict our analysis to a subset of bright GP sources by employing the following criteria: Using the unWISE full-depth source catalog \citep{Schlafly2019} that is based on significantly deeper ($2\times$) imaging than the AllWISE Source Catalog, we cross-matched it with our GP sample using a search radius of 2\arcsec\ and found 730 mid-IR counterparts. Considering the completeness and reliability estimates of the catalog from \citealt{Schlafly2019}, we select only sources with W1 $\leq$ 16 mag, W2 $\leq$ 15.7 mag for the variability analysis. After eliminating duplicate sources which are detected in multiple coadds, our final sample contains 112 sources.

\section{AGN selection}\label{sec:agn_sel}

\begin{figure}
	\epsscale{1.1}
	\plotone{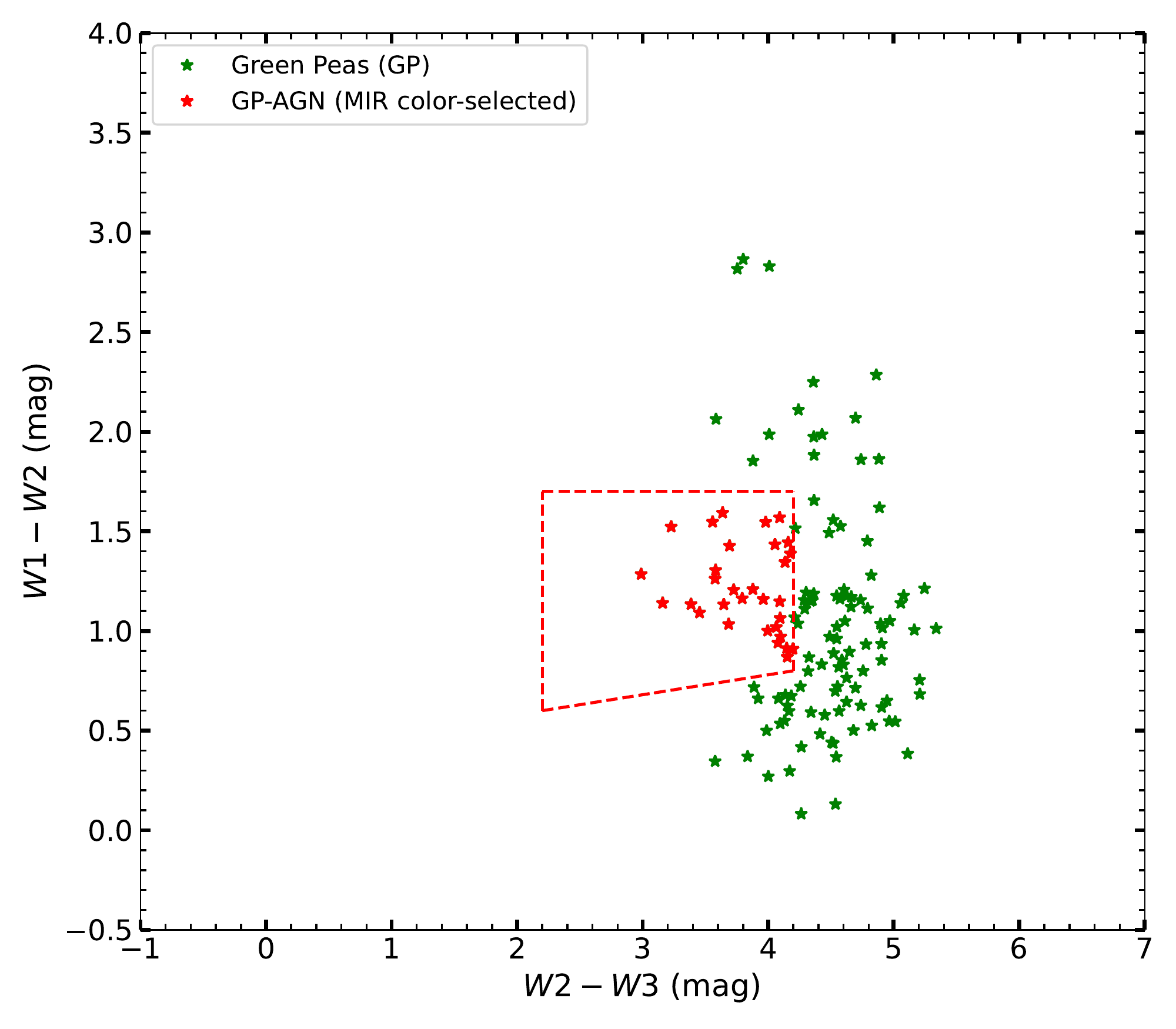}
	\caption{WISE color-color diagram for Green Peas. AGN candidates selected by the \citet{Jarrett2011} criteria are shown as \emph{red diamonds}.}
	\label{fig:midIR_Jarrett}
\end{figure}

\subsection{WISE MIR colors}\label{sec:midIR_col}
Several MIR color diagnostics exist in literature for selecting AGN, which are mostly based on data from either \emph{Spitzer} (e.g., \citealt{Stern2005, Lacy2007, Donley2012}) or \wise\ (e.g, \citealt{Jarrett2011, Stern2012,Mateos2012}). Since we are dealing with observations from \wise, we required our selection criteria to be defined based on \wise\ data that utilizes either one- or two-color criteria to select AGN. The one-color \wonetwo\ criterion, however, suffers from contamination from star-forming galaxies (e.g., \citealt{Hainline2016}). Therefore, we employed the more stringent two-color criteria by \citet{Jarrett2011} to select AGN candidates from our GP sample. As mentioned in Section \ref{sec:wise_midIR}, WISE photometry was available for 516 GPs from the AllWISE Source Catalog; however, only 266 of them had SNR $\geq 5$ in \won, \wtw, and \wth\ bands. For each of these GPs, given the large PSF size of the WISE bands, we visually inspected corresponding optical imaging from Legacy Survey (\citealt{Dey2019}; with a median 5$\sigma$ depth $\sim 22.5$ AB mag in $z$-band) for contamination within a radius of 15' and found 132 GPs with neighboring sources that exhibit blending with the GP in the relevant WISE bands. For further analysis involving the MIR colors, we consider only the 134 GPs without contamination.

The \wth\ bandpass is sensitive to PAH (11.3$\mu$m) emission from nearby galaxies and shorter wavelength PAH (6.2 and 7.7$\mu$m) emission from redshifted galaxies, as well as warm continuum from AGN at both low and high redshift.  A typical ``AGN'' region in the MIR color space, encompassing QSOs and Seyfert galaxies, defined by \citet{Jarrett2011} is as follows:

\begin{equation}
	\begin{gathered}
		2.2 < [W2- W3] < 4.2 \text{, and}\\ (0.1 \times [W2 - W3] + 0.38) < [W1 - W2] < 1.7
	\end{gathered}
\end{equation}

31 GPs were selected as AGN using these criteria. Figure \ref{fig:midIR_Jarrett} shows the location of GPs in the \wise\ color-color space and the selected AGN candidates based on this criteria.  We note that this selection will mostly include AGN that are significant brighter than host galaxy emission.  For weaker AGN, whose continuum emission is diluted by the host galaxy emission, their \wonetwo\ color can get bluer such that they become inseparable from normal star-forming galaxies \citep{Stern2012}.

\begin{figure}
	\includegraphics[scale=0.54]{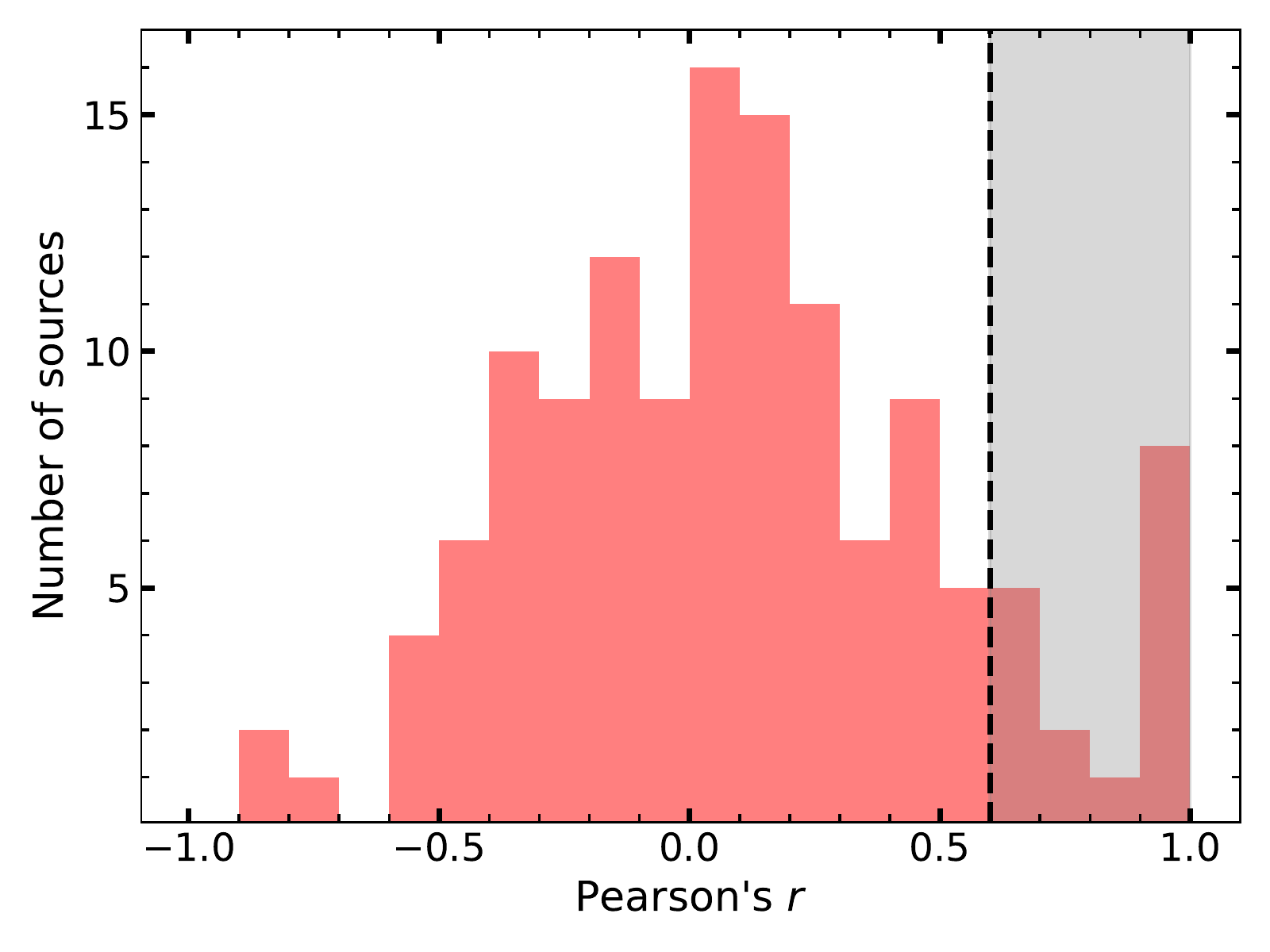}
	\caption{Distribution of Pearson correlation coefficient ($r$) of \won\ and \wtw\ for GPs with photometric observations in at least 5 epochs in \won\ and \wtw. The \emph{grey} shaded region denotes the sources with high correlation which are considered for variability selection along with other criteria, as mentioned in Section  \ref{sec:midIR_var}.}
	\label{fig:Pearson}
\end{figure}

\subsection{WISE MIR variability}\label{sec:midIR_var} 

Compared to optical variability arising from the accretion disk emission of an AGN, the MIR variability due to emission from the hot dust torus is expected to be smoothed out at shorter timescales because IR variations originate from a much bigger region than the optical variations \citep{Kozlowski2016}. Therefore, in this work, we focus on exploring the long-term variability of GPs. For the 112 GPs in our sample, we performed forced-aperture photometry on the time-resolved unWISE coadds (mentioned in Section \ref{sec:data}) and construct light curves using the following procedure.

For each GP, we determine the unWISE coadd tile containing the GP using the coadd\_id column in the source catalog from \citet{Schlafly2019}. The Epochal Coadd Index Table from \citet{Meisner2018} contains information about the available coadd epochs for a given coadd\_id in each of the bands, W1 and W2. Based on this table, we obtain the coadds for all relevant epochs for each GP with a given coadd\_id.

Given the WISE survey strategy, \citet{Meisner2018} found that subtracting pairs of consecutive coadd epochs (with same coadd\_id and WISE band) resulted in dipole residuals suggestive of astrometric misalignments. These residuals are due to the asymmetric PSF models adopted by WISE with respect to swapping the scan direction (which can be forward-pointing/back-ward pointing corresponding to the scan pointing forward/backward along the Earth’s orbit). In order to redress this issue, \citet{Meisner2018} provide “recalibrated” astrometric solutions for each epochal coadd where the WCS calibration was performed using flux-weighted centroids. Before performing photometry, we applied the WCS recalibration solutions given in the Epochal Coadd Index Table to each of the relevant coadd images.

For each GP in our sample, since the coadd images are quite large ($\sim1.5\deg \times 1.5\deg$), we generated smaller cutout images ($\sim$1.5’ $\times$ 1.5’) for all epochal coadd images, centered around the GP coordinates from SDSS. Subsequently, we performed forced-aperture photometry using Source Extractor (SExtractor; \citet{Bertin1996}) in dual-mode with a fake detection image containing a single Gaussian source whose FWHM is 6” ($\sim$PSF FWHM of the WISE bands). Our aperture has a diameter of $\sim$12” which is roughly 2 times the PSF FWHM of \won\ and \wtw. We construct the individual light curves in \won\ and \wtw\ using the measured aperture photometry from multiple epochs for each GP.

\begin{figure*}
	\begin{tabular}{ll}
		\hspace{-2.5em}\includegraphics[width=3.5in]{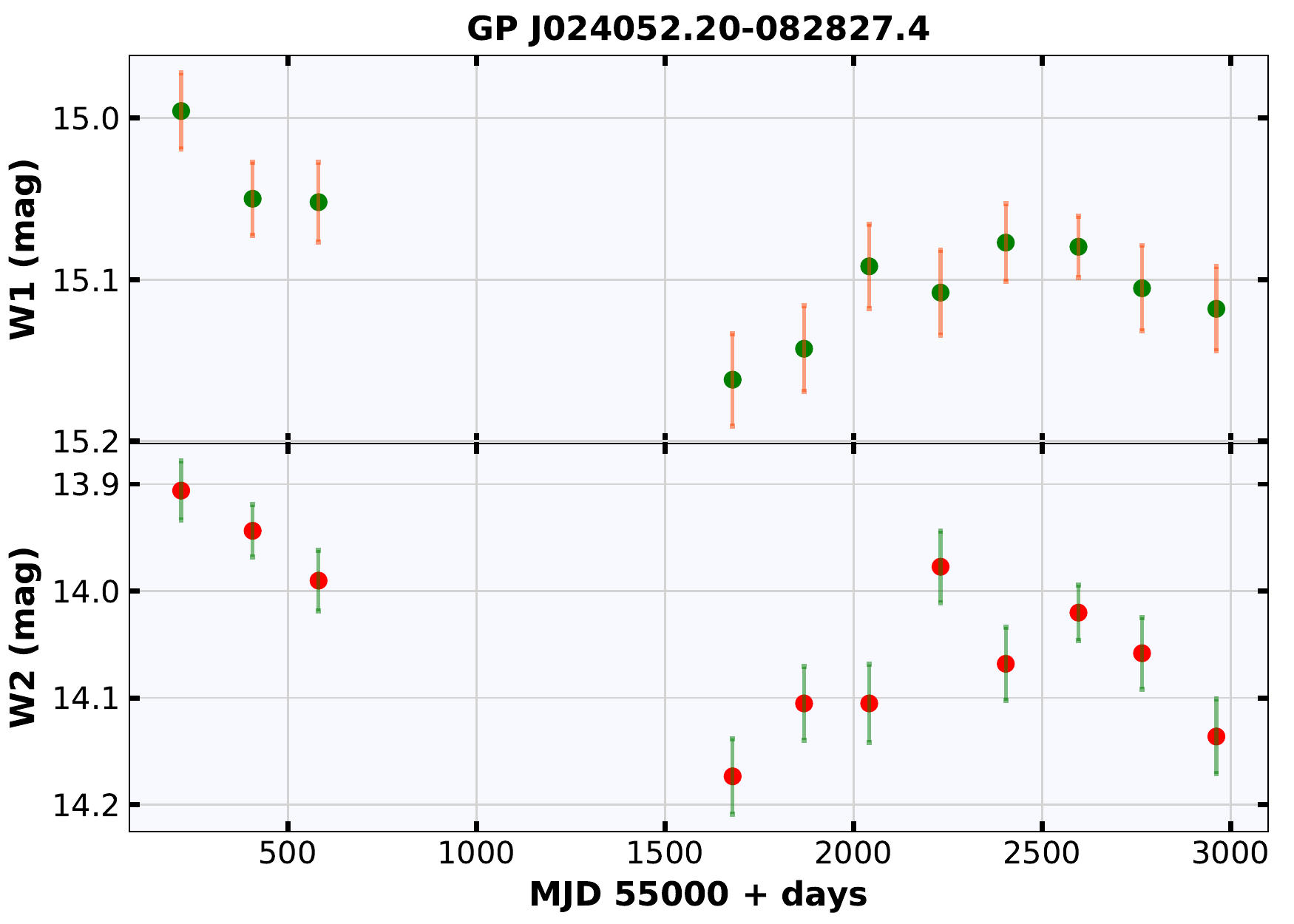} & \includegraphics[width=3.5in]{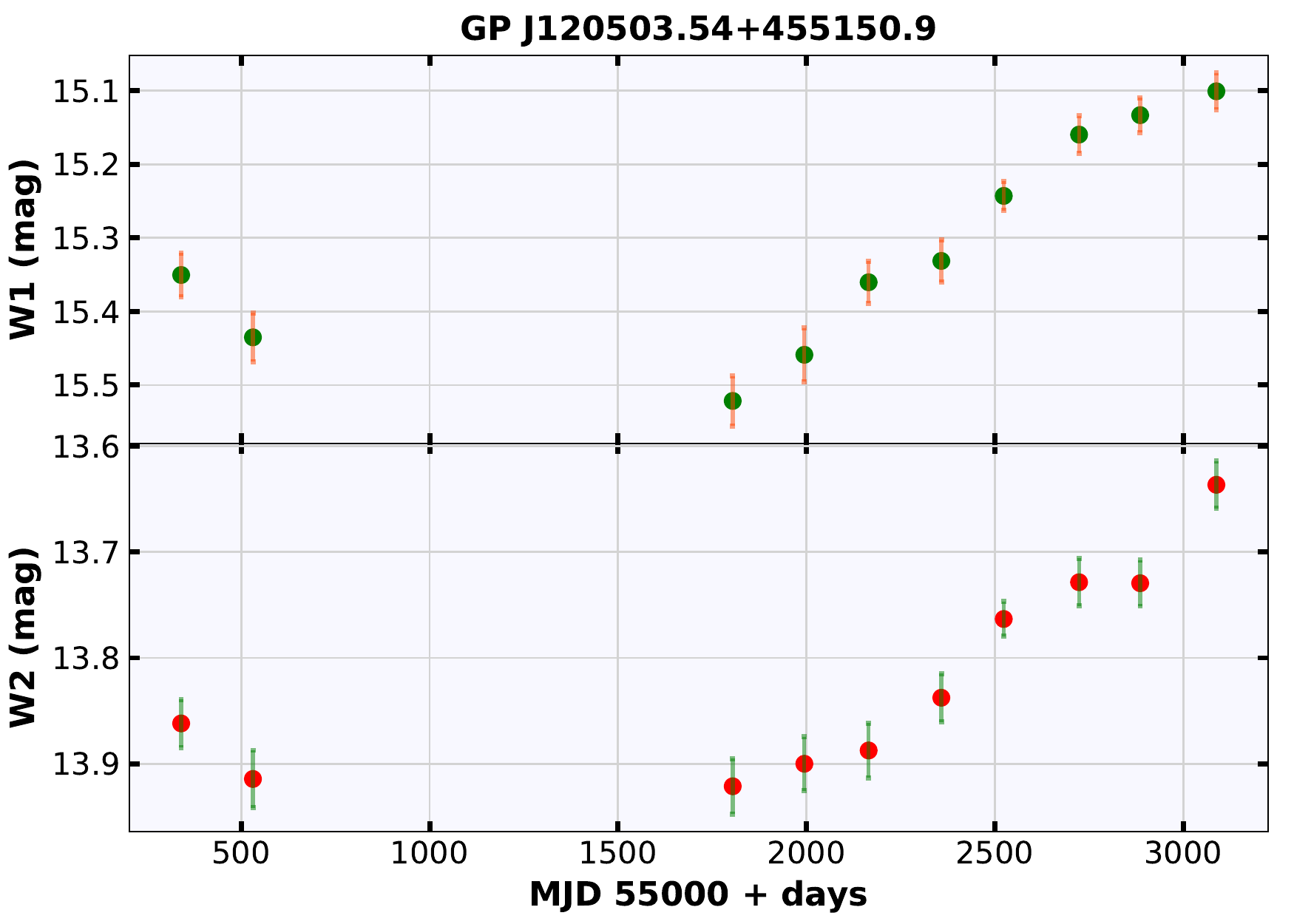} \\
	\end{tabular}
	\caption{MIR light curves in \emph{WISE} \won\ (\emph{top}) and \wtw\ bands (\emph{bottom}) for the variable GP sources. Magnitudes are based on the Vega system. The gap in measurements (between 55600 and 56700 days) corresponds to the time when \emph{WISE} was in hibernation mode.}
	\label{fig:WISE_lc}
\end{figure*}

Following \citet{Ai2010} and \citet{Kozlowski2010}, we employed the amplitude of variability ($\sigma_m$) and Pearson correlation coefficient ($r_{12}$) metrics, based on photometry from \won\ and \wtw, to select variable candidates among our GP sample. For sources with light curves in \won\ and \wtw, we calculated Pearson's $r_{12}$ as 
\begin{equation}
	r_{12} = \frac{C_{12}}{\Sigma_{W1} \Sigma_{W2}}
\end{equation}
where C$_{12}$ is the covariance between \won\ and \wtw\ bands for N-epoch (N$_{ep}$) measurements,
\begin{equation}
	\begin{aligned}
		C_{12} = \frac{1}{N_{ep}-1} \sum\limits_{i}^{N_{ep}} (m[W1]_i-\langle m[W1] \rangle) \times \\ (m[W2]_i-\langle m[W2] \rangle),
	\end{aligned}
\end{equation} 
$\Sigma_{W1}$ and $\Sigma_{W2}$ are the standard deviations in \won\ and \wtw, respectively,
\begin{equation}
	\Sigma_{W1} = \sqrt{\frac{1}{N_{ep}-1} \sum\limits_{i}^{N_{ep}} (m[W1]_i-\langle m[W1] \rangle)^2},
	\label{eq:sigdef1}
\end{equation}
\begin{equation}
	\Sigma_{W2} = \sqrt{\frac{1}{N_{ep}-1} \sum\limits_{i}^{N_{ep}} (m[W2]_i-\langle m[W2] \rangle)^2},
	\label{eq:sigdef2}
\end{equation}
where $m[W1]_i$, $m[W2]_i$ are the \won, \wtw\ magnitudes in the $i$th epoch, and $\langle m[W1] \rangle$, $\langle m[W2] \rangle$ are the average magnitudes in \won, \wtw\ respectively, for a given GP source. The amplitude of variability, $\sigma_m$ is derived as,
\begin{equation}
	\sigma_m =
	\begin{cases}
		\sqrt{\Sigma^2 - \epsilon^2}, & \quad \text{if $\Sigma > \epsilon$} \\ 0, & \quad \text{otherwise}
	\end{cases}
\end{equation}
where $\Sigma$ is the photometric variance as defined in equations~\ref{eq:sigdef1}--\ref{eq:sigdef2}, and $\epsilon$ is the variance due to measurement errors for $N$ observations in a given band and is defined as,
\begin{equation} \label{eqn:eps}
	\epsilon^2 = \frac{1}{N} \sum\limits_{i}^{N} \epsilon^2_i + \epsilon^2_s
\end{equation}
where $\epsilon_i$ is the photometric error of $i$th epoch, based on the scatter among individual measurements within that epoch, and $\epsilon_s$ is the systematic error. For \won\ and \wtw, the reported systematic uncertainties are 0.024 mag and 0.028 mag, respectively \citep{Jarrett2011}, which are added in quadrature to the photometric uncertainties ($\epsilon_s$ in Eqn.\ \ref{eqn:eps}). 

True variable objects are those with positive variability amplitudes ($\sigma_{W1}$, $\sigma_{W2} > 0$) and strong correlation between \won\ and \wtw. Four GP sources were identified as potential AGN candidates fulfilling the following criteria: (1) N$_{ep} \geqslant 5$, (2) $\sigma_1$, $\sigma_2 > 0$, and (3) $r_{12} > 0.6$. For these sources, we examined the corresponding optical image in SDSS for possible contaminants/blending and found two of these to be affected by neighboring sources within 15” from the GP coordinates.

In addition to the above, we measured the $\chi^2$ in both bands (\won\ and \wtw) relative to the best-fitting flux, to test the likelihood of the data under the null hypothesis of a constant source. We found high $\chi^2$ and low probabilities (p $<0.01$) of the null hypothesis in three of the four GPs. With a p-value of $>0.1$ in both bands, we are unable to reject the possibility of GP J143202.85+515252.2 being a constant source. We also performed a simple Monte Carlo simulation to determine whether the observed variability amplitudes ($\sigma_{W1}$ and $\sigma_{W2}$) are obtained by chance. For this simulation, we construct the non-varying light curves for each GP as follows: Using the mean GP flux and the corresponding observed errors in each epoch, we generate fake WISE photometry across all epochs in both W1 and W2 bands and measure their variability amplitudes. We repeat this procedure for 100,000 realizations. Based on the observed distribution, we calculated the probability of randomly obtaining a variability amplitude (in each band) that is equal to or greater than the observed value and found them to be highly insignificant ($\lesssim$ 1\%) except for the GP J143202.85+515252.2.

Eventually, excluding J143202.85+515252.2, we conclude that there are three GP objects that are MIR variable in W1 and W2. Light-curves of GP J024052.20-082827.4 and J120503.54+455150.9 are shown in Figure \ref{fig:WISE_lc} and their properties are given in Table \ref{tab:gp_props}. Additionally, GP J132738.16+132444.5 passed our tests of variability but is among the sources whose WISE photometry is affected by close neighbors.

Recently, \cite{Secrest2020} performed a search for variable AGN in the MIR using \mbox{\wise/\neowise\ }for a sample of dwarf galaxies and found significant variability in 0.09\% of their sample. Another recent study \citep{Ward2021} searched for intermediate mass BH candidates based on variability, using \mbox{\wise\ }, in a bigger dwarf galaxy sample and found 0.2\% of their sample to contain variable AGN. Compared to these previous studies, we find a higher fraction ($\sim 1.98\%$) of variable AGN in our GP sample.  The Poisson probability of finding 2 variables among 112 sources would be only $\sim 0.5\%$ and $\sim 2\%$, assuming the variability fractions found by \cite{Secrest2020} and \cite{Ward2021}, respectively. This suggests that GPs host AGN more often than the parent samples of those two studies.

\begin{figure}
	\epsscale{1.2}
	\plotone{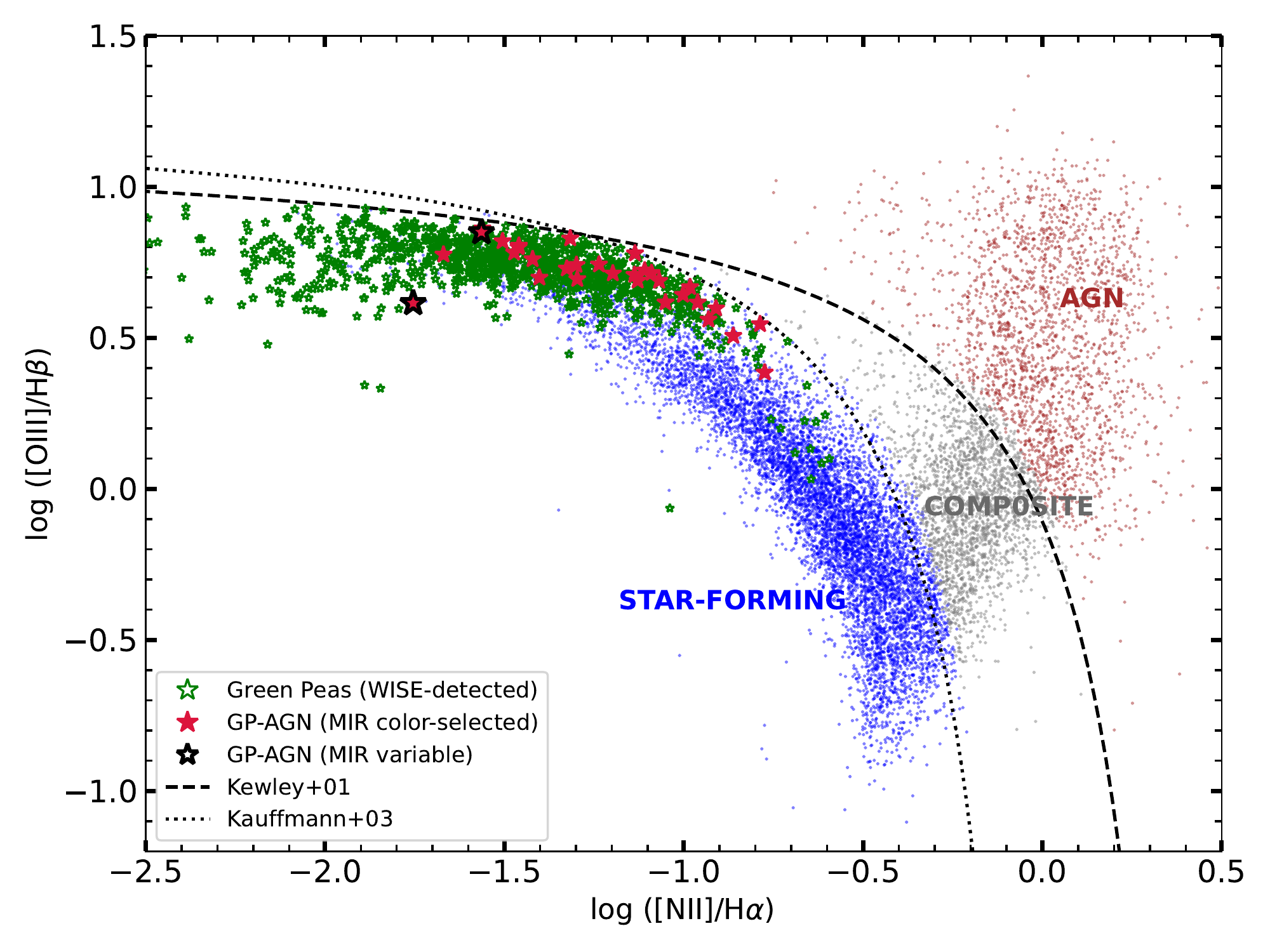}
	\caption{GPs on the BPT diagram. Shown for comparison are SDSS star-forming galaxies (\emph{blue}), composite galaxies (\emph{grey}) and AGN (\emph{red}) from the MPA-JHU catalog . The demarcation lines for star-forming galaxy and AGN populations from \citet{Kewley2001} is represented by \emph{dashed-line} and \cite{Kauffmann2003} is represented by \emph{dotted-line}. GPs selected as AGN based on MIR variability (\emph{black}) and MIR color-color criteria (\citealt{Jarrett2011}; \emph{red}) are also shown.}
	\label{fig:bpt}
\end{figure}

\section{Discussion} \label{sec:disc}

\subsection{Limitations of BPT selection} \label{sec:bpt_cons}
The classical BPT diagram \citep{Baldwin1981}, log([N\textsc{ii}]/H$\alpha$) vs. log([O\textsc{iii}]/H$\beta$), is a widely employed diagnostic to differentiate AGN from galaxies. Most strong AGN with solar/super-solar metallicity tend to occupy the ``rising branch" on the right-hand side of the diagram (Figure \ref{fig:bpt}).    However, low metallicity (sub-solar or less) AGN occupy a region on the left-edge of the diagram where they cannot be easily distinguished from star-forming galaxies. Based on chemical evolution estimates from cosmological hydrodynamic simulations, \citealp{Kewley2013} showed that low-metallicity galaxies (12+log(O/H) $\lesssim$ 8.4) containing AGN will be impossible to distinguish from low-metallicity star-forming galaxies using the BPT diagram. $\sim$98\% of GP objects have metallicities below this limit.  Therefore, the BPT diagram is inefficient at selecting AGN candidates among the GP sample. 

\begin{figure}
	\epsscale{1.2}
	\plotone{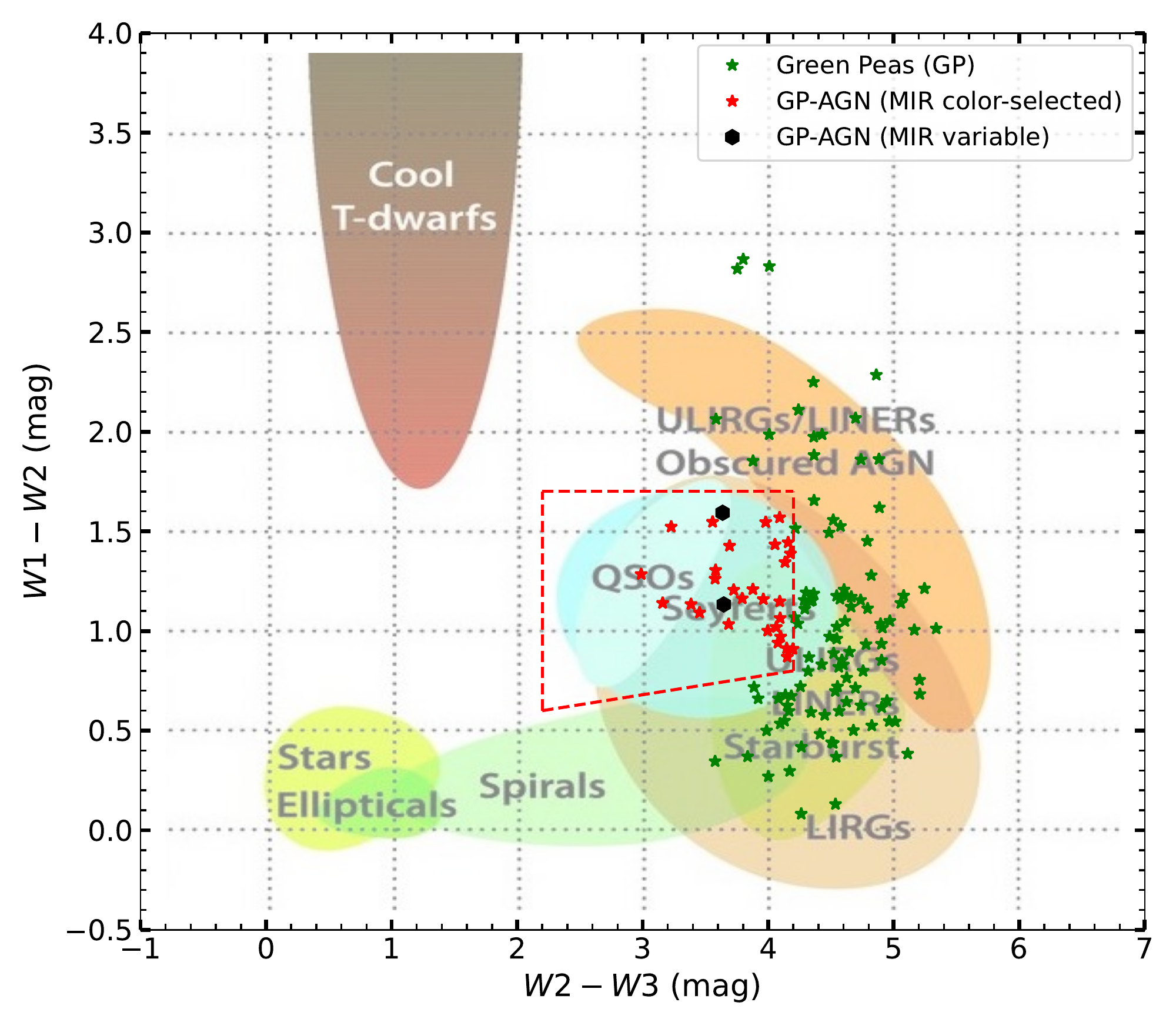}
	\caption{WISE color-color diagram for GP candidate AGN selected based on MIR variability (\emph{black}). For comparison, the color-selected AGN based on \citet{Jarrett2011} criteria are also shown (\emph{red}).}
	\label{fig:midIR_variable}
\end{figure}

\begin{deluxetable*}{ccccccc}
	\tablenum{1}
	\tablecaption{Properties of Green Peas exhibiting variability in the MIR \label{tab:gp_props}}
	\tablewidth{0pt}
	\tablehead{
		\colhead{ID} & \colhead{R. A.} & \colhead{Dec.} & \colhead{$z_{spec}$} &  \colhead{$\sigma_{W1}$} & \colhead{$\sigma_{W2}$} & \colhead{Pearson \emph{r}}  \\
		\colhead{} & \colhead{(J2000)} & \colhead{(J2000)} & \colhead{}  & \colhead{(mag)} & \colhead{(mag)} & \colhead{}
	}
	\startdata
	J024052.20-082827.4 & 02:40:52.20 & -08:28:27.4 & 0.0822&0.03 & 0.07 & 0.85\\
	J120503.54+455150.9 & 12:05:03.54 & +45:51:50.9 & 0.0654& 0.14 & 0.09 & 0.96\\
	\enddata
	\tablecomments{Column 1: Green Pea ID, column 2: Right Ascension (J2000), column 3: Declination (J2000), column 4: spectroscopic redshift from SDSS, column 5: variability amplitude in \won, column 6: variability amplitude in \wtw, column 7: Pearson \emph{r} correlation coefficient for \won\ and \wtw.}
\end{deluxetable*}

\subsection{MIR color-selected AGN candidates}
Different methods of AGN selection can be sensitive to different type of AGN with varying luminosities, redshifts, line-of-sight angles, obscuration, and so on.  Although  X-ray detection is considered to be the most robust diagnostic of AGN presence, a significant fraction of AGN that are selected using infrared diagnostics are not detected in X-rays (e.g., \citealt{Simmonds2016}). Such sources are expected to be significantly obscured and/or low-luminosity AGN.

MIR color-based selection relies on the fact that hard radiation from an AGN heats the surrounding dusty torus to high temperatures, limited only by the grain sublimation temperature. This produces an approximate power-law continuum in the MIR that is easily distinguishable from stars and typical star-forming galaxies (e.g., \citealt{Stern2012,Assef2013}). Since MIR is insensitive to extinction, this selection is able to identify AGN, especially obscured ones that are optically hidden (e.g., \citealt{Assef2013,Stern2014}).

Even though red MIR colors are a strong indication of AGN activity in such galaxies, some studies in the past have shown that intense star formation activity in low-metallicity starburst galaxies (such as blue compact dwarfs (BCDs), which are similar to GPs in many ways) can heat dust to high temperatures that produce extremely red MIR colors (e.g., \citealt{Griffith2011,Izotov2014}). Considering that single-color criteria can suffer from contamination due to star-forming galaxies (e.g., \citealt{Hainline2016}), we employed the two-color criteria for selection of AGN candidates which is more reliable and robust. Also, our GP sample has a less extreme typical metallicity (12+log (O/H) $\sim$ 8.05; \citealt{Jiang2019}), compared to the BCDs in \citealt{Griffith2011} (12+log (O/H) $\sim$ 7.46) and \citealt{Izotov2014} (12+log (O/H) $\sim$ 7.76), suggesting that GP stellar populations are less likely to produce the observed hot dust emission in the MIR. In fact, our color- and variability-selected AGN candidate sample has a median 12+log (O/H) $\sim$ 8.17. Also, according to \cite{Satyapal2018}, low-metallicity galaxies with low dust abundance require high column densities, $N_H$, to produce extreme red WISE colors, while GPs are observed to have low column densities \citep[$N_H \lesssim 10^{20}$ cm$^{-2}$; e.g.,][]{Yang2016,Yang2017b}. 

Given an accuracy of $\sim$0.1 mag, and a time scale of 300-3000 days for flux variation, we expect $\sim$30\%-50\% of AGN to show detectable variation based on the observed quasar variability estimates from \cite{MacLeod2012}. Using the \emph{Spitzer} Deep Wide-Field Survey, \cite{Kozlowski2010} found that $\sim$15\% of AGN in their sample were sufficiently variable to be selected as a variable AGN. The two AGN observed to vary in our sample then imply $\sim 4$--$15$ AGN in total.
This is broadly consistent with \cite{Assef2013}, who found $>75\%$ reliabilty for mid-IR selection of AGN with the criteria of \cite{Jarrett2011} in a  brightness range similar to our GP sample ($W2 < 15.73$). 

\subsection{MIR variable AGN candidates}

\begin{figure}
	\epsscale{1.2}
	\plotone{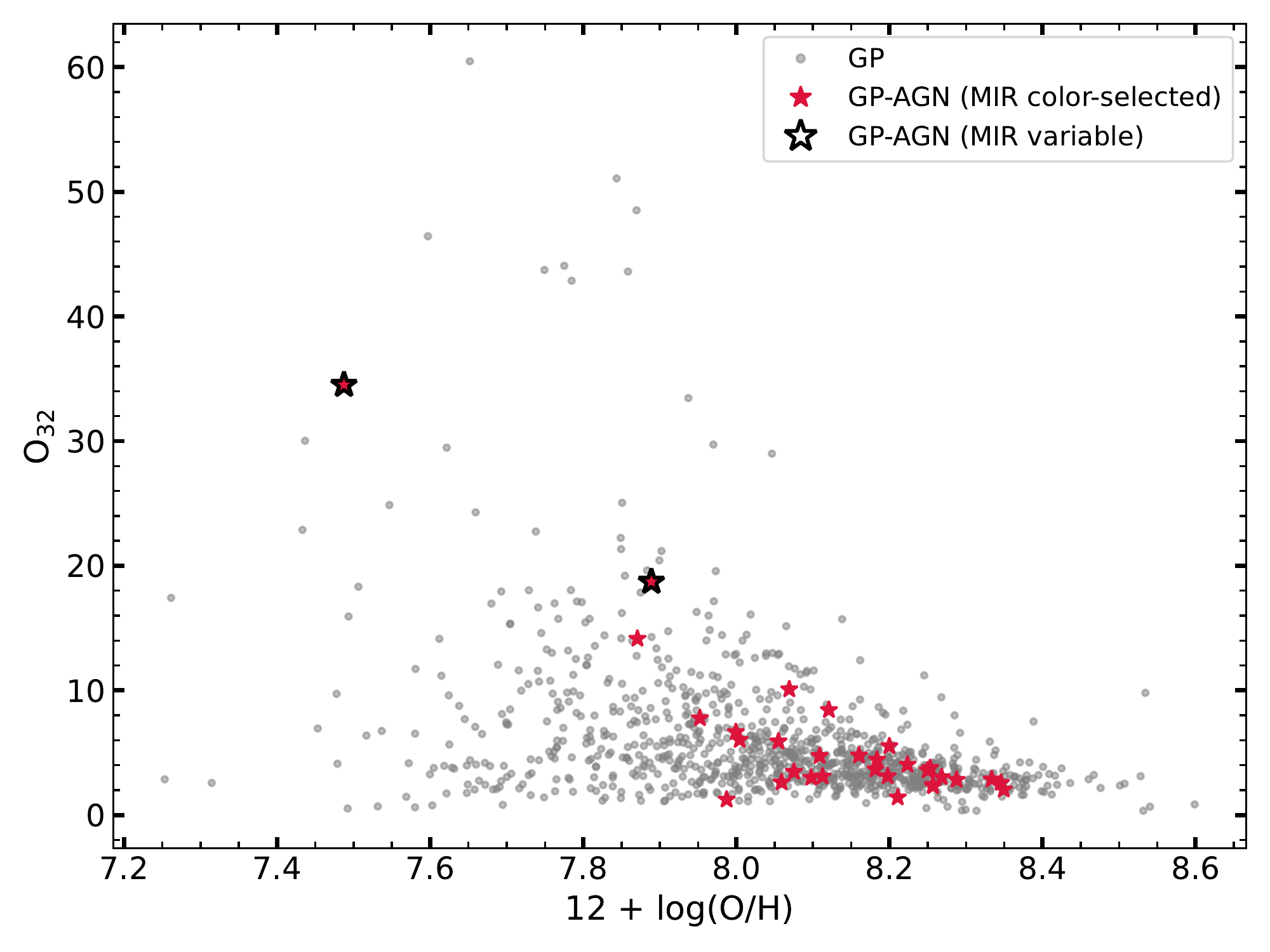}
	\caption{Gas-phase metallicity vs.\  extinction-corrected [\textsc{Oiii}]/[\textsc{Oii}] (O$_{32}$) for GPs. The low metallicity and high O$_{32}$ candidate AGN selected based on MIR variability (\emph{black}) are shown along with the MIR color-selected AGN candidates (\emph{red}).}
	\label{fig:metallicity_O32}
\end{figure}

Both variable AGN candidates identified in this work are also identified as AGN by MIR color selection (Figure \ref{fig:midIR_variable}). Considering the location of various classes of objects in the WISE color-color space (as shown in Figure 12 from \cite{Wright2010}), all our AGN candidates exhibit colors that coincide with colors typically observed among QSOs/Seyfert objects. Of all the AGN candidates, the variable GPs have the highest [\textsc{Oiii}]/[\textsc{Oii}] ratio (O$_{32}>18$) and are also among the most metal-poor objects with gas-phase metallicities, 12+log (O/H) $\lesssim$ 7.9, as shown in Figure \ref{fig:metallicity_O32}.

In addition to the MIR AGN signatures, several other characteristics indicative of AGN have been observed in the optical spectra of these two variable GPs. Recently, \cite{Izotov2017,Izotov2021} have detected high-ionization emission lines such as [Ne \textsc{v}] $\lambda3426$ and He \textsc{ii} $\lambda4686$, indicating the presence of hard ionizing radiation in these galaxies. In order to determine the physical processes responsible for the hard radiation, they compare observations with models of photoionized H \textsc{ii} regions using CLOUDY \citep{Ferland2013}, in combination with (1) ionizing radiation derived from BPASS stellar population models \citep{Eldridge2017} as well as a combination of stellar radiation from STARBURST99 models \citep{Leitherer1999} with ionizing radiation produced by shocks or by non-thermal radiation from an AGN. They find that pure stellar ionizing radiation is unlikely to produce the high-ionization emission in those galaxies and attribute the hard radiation to fast radiative shocks with velocities up to $\sim 500$ km s$^{-1}$. However, they also conclude that their observations are also consistent with the presence of an AGN source, contributing $\sim$10-20 percent to the total luminosity of ionizing radiation in all those galaxies. In addition to the high ionization lines, they also find low-intensity broad components of H$\alpha$ in the optical spectra for both these variable GPs and this could arise from the gas accretion onto a black hole in an AGN. Recently, \cite{Burke2021} also found persistent broad Balmer emission lines in three metal-poor dwarf emission-line galaxies ($z \lesssim 0.3$) with AGN-like variability in the optical.

Given the observed variability and redder colors in the MIR and the aforementioned optical characteristics, we conclude that these two variable GPs most likely contain a low-mass, low-luminosity AGN, with gas-phase metallicities lower than that of a typical AGN.

\section{Conclusions}\label{sec:concl}

Using a sample of $\sim$1000 GPs selected from SDSS, based on strong [O\textsc{iii}]$_{\lambda5007}$ and/or H$\beta$ emission, we searched for AGN presence based on MIR colors and variability using \emph{WISE/NEOWISE-R} observations. Considering only $>5\sigma$ detections in the shorter three \emph{WISE} bands with no contamination from neighboring sources, 134 GPs were used for AGN selection based on MIR colors. For variability-based AGN selection, 112 GPs with \emph{W}1 $\leqslant 16$ and no contamination from neighboring sources were used. Our main findings are as follows: 
\renewcommand{\theenumi}{\roman{enumi}}
\begin{enumerate}
	\item Based on photometry from the shortest three \emph{WISE} bands, we selected a sample of 31 AGN (out of 134 GPs) candidates using the color-color criteria from \cite{Jarrett2011}. 
	
	\item Using time-resolved unWISE coadd images, we derive multi-year light curves in \won\ and \wtw\ bands for a subset of bright GPs in the mid-infrared. Based on variability amplitude and Pearson's correlation coefficient ($r$) metrics, we identified 2 GPs (out of 112) that exhibit notable variability in the mid-infrared. 
	
	\item Both variable GPs are selected as AGN also based on MIR colors. In addition to the MIR signatures, optical spectra of these variable GPs from previous studies \mbox{\citep{Izotov2017,Izotov2021}} show evidence of hard ionizing radiation based on the presence of high ionization lines such as [Ne \textsc{v}] and He \textsc{ii}. They also find low-intensity broad component emission in the H$\alpha$ line in these GPs.  Given the variability we observe, these optical signatures are best interpreted as additional indications of gas accretion onto a central BH.
\end{enumerate}

Our findings suggest MIR colors and variability can select AGN candidates in low-mass galaxies that elude detection via other selection methods. Since GPs are extreme emission-line galaxies with high specific SFRs and low typical metallicities, optical selection methods such as those based on emission-line ratios may be unable to detect the presence of AGN in GPs, both because optical emission lines may be dominated by star formation activity, and because the optical line ratio signature of AGN becomes less distinct from star formation at low metallicity.
Multi-wavelength follow-up observations of these sources will be crucial in validating the AGN presence and constraining their properties, including BH masses and accretion rates. If any of these GPs are shown to not host an AGN, then investigation regarding the origin of hot dust and flux variability will be of great interest in itself.

\acknowledgments

We thank an anonymous referee and Nathan Secrest for their helpful comments and suggestions on an earlier draft, which significantly improved the manuscript. This publication makes use of data products from the Wide-field Infrared Survey Explorer, which is a joint project of the University of California, Los Angeles, and the Jet Propulsion Laboratory/California Institute of Technology, and NEOWISE, which is a project of the Jet Propulsion Laboratory/California Institute of Technology. WISE and NEOWISE are funded by the National Aeronautics and Space Administration. 

Funding for the Sloan Digital Sky Survey IV has been provided by the Alfred P. Sloan Foundation, the U.S. Department of Energy Office of Science, and the Participating Institutions. SDSS-IV acknowledges support and resources from the Center for High Performance Computing  at the University of Utah. The SDSS website is www.sdss.org.

SDSS-IV is managed by the Astrophysical Research Consortium for the Participating Institutions of the SDSS Collaboration including the Brazilian Participation Group, the Carnegie Institution for Science, Carnegie Mellon University, Center for Astrophysics | Harvard \& Smithsonian, the Chilean Participation Group, the French Participation Group, Instituto de Astrof\'isica de Canarias, The Johns Hopkins University, Kavli Institute for the Physics and Mathematics of the Universe (IPMU) / University of Tokyo, the Korean Participation Group, Lawrence Berkeley National Laboratory, Leibniz Institut f\"ur Astrophysik Potsdam (AIP),  Max-Planck-Institut f\"ur Astronomie (MPIA Heidelberg), Max-Planck-Institut f\"ur Astrophysik (MPA Garching), Max-Planck-Institut f\"ur Extraterrestrische Physik (MPE), National Astronomical Observatories of China, New Mexico State University, New York University, University of Notre Dame, Observat\'ario Nacional / MCTI, The Ohio State University, Pennsylvania State University, Shanghai Astronomical Observatory, United Kingdom Participation Group, Universidad Nacional Aut\'onoma de M\'exico, University of Arizona, University of Colorado Boulder, University of Oxford, University of Portsmouth, University of Utah, University of Virginia, University of Washington, University of Wisconsin, Vanderbilt University, and Yale University.

This work has made use of the following open-source softwares: SciPy \citep{Virtanen2020}, NumPy \citep{Harris2020}, Matplotlib \citep{Hunter2007}, Astropy \citep{Astropy2013,Astropy2018}, DS9 \citep{Joye2003} and TOPCAT \citep{Taylor2005}.

\bibliography{refs}{}
\bibliographystyle{aasjournal}



\end{document}